# Magnetization reversal in Py/Gd heterostructures

Pavel N. Lapa,[1,2] Junjia Ding,[1] John E. Pearson,[1] Valentine Novosad,[1] J. S. Jiang,[1] and Axel Hoffmann[1]

[1]*Materials Science Division, Argonne National Laboratory, Argonne, Illinois 60439, USA*

[2]*Department of Physics and Astronomy, Texas A&M University, College Station, Texas 77843-4242, USA*

Using a combination of magnetometry and magnetotransport techniques, we studied temperature and magnetic field behavior of magnetization in Py/Gd heterostructures. It was shown quantitatively that proximity with Py enhances magnetic order of Gd. Micromagnetic simulations demonstrate that a spin-flop transition observed in a Py/Gd bilayer is due to exchange-spring rotation of magnetization in the Gd layer. Transport measurements show that the magnetoresistance of a [Py(2 nm)/Gd(2 nm)]$_{25}$ multilayer changes sign at the compensation temperature and below 20 K. The positive magnetoresistance above the compensation temperature can be attributed to an in-plane domain-wall, which appears because of the structural inhomogeneity of the film over its thickness. By measuring the angular dependence of resistance we are able to determine the angle between magnetizations in the multilayer and the magnetic field at different temperatures. The measurement reveals that due to a change of the chemical thickness profile, a non-collinear magnetization configuration is only stable in magnetic fields above 10 kOe.

## I. INTRODUCTION

The number of unique phenomena makes surface magnetism of rare-earth Gd an extremely interesting, while immensely challenging, scientific topic. It was shown that even for a Gd film grown on a non-magnetic W substrate, magnetic order near the surface can be significantly enhanced [1-3]. When Gd is in proximity with a transition metal (TM), the effect becomes even more prominent [4, 5]. For a Fe/Gd multilayer, it was demonstrated that a few atomic monolayers of Gd adjacent to the Fe layer have Curie temperature comparable to that of Fe [6], regardless of the Curie temperature of bulk Gd being only 293 K. Another interesting property of Gd/TM heterostructures is an antiferromagnetic exchange interaction between Gd and TM. This results in magnetizations of adjacent Gd and TM layers being antiparallel to each other without any magnetic fields applied. The unusual antiferromagnetic coupling in combination with the enhancement of magnetic order triggered the use of Gd/TM heterostructures for artificial ferrimagnets applications. As of today, artificial ferrimagnets have been implemented for spin mixing in superconducting spin valves [7] and an out-of-plane anisotropy in Gd/Co [8] and Gd/Fe [9, 10] multilayers stimulates the study of the artificial ferrimagnets for a bubble domain application [11-13].

Furthermore, due to the lower ordering temperature of Gd compared to many ferromagnetic TM materials, the magnetization in the Gd layers demonstrates a stronger temperature dependence than the magnetization in the TM layers. For certain Gd/TM heterostructures, at a so-called "compensation temperature", the magnetic moment of the Gd layer becomes equal to the magnetic moment of the TM layer. Because of the antiferromagnetic coupling between the layers, this compensation results in a vanishing total magnetization for these Gd/TM multilayers. Therefore, it is of particular interest to determine magnetization configurations and magnetization reversal mechanisms at the compensation temperature, where magnetic moments of the Gd and TM layers become equal. From a fundamental perspective, Camley and Tilley, using mean-field calculations, showed that, a so-called twisted magnetic state with non-collinear configuration of magnetization over a film's thickness can appear at the compensation temperature [14]. That is the magnetization of layers located near the top and bottom surfaces of a film begin to cant, resulting in a well-defined angle with respect to the external magnetic field. As the temperature approaches the compensation temperature, the twists propagate deeper into the multilayer. Camley and Tilley demonstrated that the characteristics of this twisted state strongly depends on the layered structure and its microscopic parameters. The magnetic behavior at the compensation temperature is also interesting due to potential applications. Recently, it was demonstrated that due to a thermally induced excitation, magnetization of GdFeCo films [15-18] and Gd/Fe [19] artificial ferrimagnets can be switched across the compensation temperature optically, without applying external magnetic field. It is believed that the phenomenon can be used for developing new magnetic recording media [20]. In addition, the antiferromagnetic coupling can result in the formation of interfacial domain-walls, which only form and persist for applied magnetic fields exceeding a critical field [21, 22].

The anisotropy of polycrystalline Gd and TM [23] causes a hysteretic behavior of magnetization, which drastically complicates analysis of a magnetization reversal in the Gd/TM heterostructures. This complication can be minimized by using Permalloy (Py = $Ni_{0.81}Fe_{0.19}$), which is a transition metal alloy with very low anisotropy. Thus Gd and Py are very promising materials for fabrication of artificial ferrimagnets. Ranchal *et al.* demonstrated that coupling between Gd and Py is antiferromagnetic [24, 25], and, important for future discussion, the intermixing of these materials strongly reduces their coupling energy [26]. Our motivation was to characterize quantitatively the coupling of Gd with Py and magnetic order in these layers, as well as to investigate proximity effects in Py/Gd artificial ferrimagnets.

## II. DETAILS OF THE EXPERIMENT AND SIMULATION

Magnetron sputtering at ambient temperatures was used to fabricate three groups of films for the study. All samples were grown on top of $Si/SiO_2$ substrates at room temperature. 5-nm thick Ta layers at the bottom and on top of all the films were used as a seed layer and for capping, respectively. The first group of samples consists of Py(50 nm)/Gd(4 nm) and Py(50 nm)/Au(0.5 nm)/Gd(4 nm) films. These films are designed to determine how proximity with Py influences the magnetization and exchange stiffness of Gd, as well as to evaluate the interlayer coupling between the two metals. The bilayer with a thin Au buffer serves to reveal how interlayer diffusion between Gd and Py affects the strength of the coupling and to show if the coupling can be controlled by placing a thin buffer between Py and Gd. The second group consists of Py/Gd multilayers [Py($t$)/Gd($t$)]$_{25}$, where $t$ is 1 or 2 nm. The purpose of studying these samples is to determine the mechanism of the magnetization reversal in the vicinity of the compensation temperature. Finally, the third group of samples is designed for estimating the effective exchange stiffness in the Py/Gd multilayers. These samples are composed of a thick Py layer adjacent to a Py/Gd stack, *i.e.*, Py(50 nm)/[Py(*t*)/Gd(*t*)]$_{25}$, *t* =1 or 2 nm. Magnetic moments were measured using a Quantum Design SQUID magnetometer. Transport measurements were conducted using a conventional four-probe technique; the films were cut into shape of $9 \times 2$ mm$^2$ stripes. Magnetic field was applied parallel to the films surface. To measure angular dependencies of resistance the stripes were installed on a horizontal rotator. The zero angle corresponds to the direction of the applied current being parallel to the external magnetic field.

In order to determine the magnetization reversal mechanism and quantitatively evaluate parameters of the Py/Gd films (exchange stiffness, magnetization, and interlayer coupling), the experimental data was simulated using the OOMMF micromagnetic simulation software [27]. Essentially, the micromagnetic model mimics a one-dimensional spin chain directed along the thickness of a film. To imitate an infinite plane, the demagnetization energy term was excluded from the calculation, while magnetization was forced to rotate in plane. Thus, these simulations yield a configuration of spins along the thickness of the films, or rephrasing, they model in-plane domain-walls occurring in the films in different magnetic fields. The size of the calculation cell in the direction perpendicular to the film plane was chosen to be 0.25 nm.

## III. BILAYER FILMS

To determine the interfacial coupling between Py and Gd and to study how proximity affects magnetic properties of the metals, we prepared the Py(50 nm)/Gd(4 nm) and Py(50 nm)/Au(0.5 nm)/Gd(4 nm) films. The idea behind the design is as follows. Keeping the Gd layer thin allows minimizing anisotropy energy term, which, in turn, enables identifying effects of the interfacial interaction more clearly. At the same time, for a relatively thick Py layer, its magnetization is predominately along the externally applied magnetic field, thus, making the Py layer effectively a "magnetic anchor". The temperature and magnetic field dependencies of magnetic moment for the films are compared with a reference Py(50 nm) film.

The temperature dependencies of the magnetic moment normalized by the film's area [Fig. 1(a)] were measured in a small magnetic field (100 Oe) applied parallel to the surface plane. It is seen that in contrast to a typical rise of magnetization with decreasing temperature as seen in the reference Py film, the total magnetic moment of the Py/Gd films (with and without Au) begins to decrease at temperatures below 120 K. This proves that the exchange interaction between Py and Gd is antiferromagnetic. Since the Gd magnetic moment developed at low temperature is antiferromagnetically coupled with the one in Py, it yields a decrease of the total magnetic moment. Importantly, the magnetic moments of the Py/Gd samples are lower than the magnetic moment of the reference Py(50 nm) film even at temperatures above 120 K. We assume that, due to proximity, a part of the Gd layer immediately adjacent to Py has a Curie temperature higher that the rest of the Gd layer. This part of the Gd layer remains ferromagnetic at higher temperatures, which reduces the magnetic moment of the Py/Gd films in comparison with the reference single-layer Py film even at temperatures above 120 K.

Magnetization curves of the Py(50 nm)/Gd(4 nm) and Py(50 nm)/Au(0.5 nm)/Gd(4 nm) films [Fig. 1(b)] demonstrate another interesting effect previously reported for Gd/Fe [28-30], Gd/Ni [4], and Gd/Co [31, 32] films, which is usually called a "spin-flop transition". Namely, at a critical field, $H_{CR} = 1.3$ kOe for the Py(50 nm)/Gd(4 nm) film, the magnetic moment exhibits an abrupt growth. Basically, in order to minimize the Zeeman energy, it becomes energetically favorable for the magnetization in the Gd layer to rotate so that it becomes aligned along the magnetic field. The microscopic mechanism of this rotation can be quite complicated, and it strongly depends on microscopic parameters of the metals. First, let us assume that the exchange stiffness of Gd and Py is high (rigid-spin approximation) and the interface coupling is comparatively weak. In this case, the spin-flop is realized by a coherent rotation of the magnetization in the entire Gd layer with respect to magnetization in Py. Micromagnetic simulations show that, under the assumption of rigid spins, the total magnetization would demonstrate a linear rise with magnetic field above $H_{CR}$ which is followed by a saturation. However, the curves in Fig. 1(b) demonstrate a different behavior; the magnetization grows nonlinearly, and the full saturation is not achieved even in high magnetic fields (40 kOe). It means that the assumption about the "weak" interface is unjustified for the system, and the switching at $H_{CR}$ does not happen due to the coherent rotation of Py and Gd magnetization with respect to each other. Another important feature of the magnetization curves is that the rigid-spin approximation fails to explain the almost linear rise of the magnetization with magnetic fields below $H_{CR}$ [inset in Fig. 1(b)].

It is clear that finite values of the exchange stiffness of Gd and Py must be taken into account for an adequate modeling of the magnetic reversal in the Py/Gd bilayer. Importantly, microscopic properties of materials at the Py/Gd interface are defined by two contrary processes. First, because the materials have different Curie temperatures, proximity is responsible for the reduction of the exchange stiffness in a thin Py sublayer adjacent to the interface, $A_{Py\_Int}$, and the enhancement of that in a thin Gd sublayer adjacent to the interface, $A_{Gd\_Int}$. Contrary to that, intermixing of Py and Gd [4, 33] results in adding Ni to Gd which strongly reduces the Curie temperature of the latter [26, 34]. This, in turn, leads to the suppression of $A_{Gd\_Int}$ and $A_{Py\_Int}$. Based on temperature and magnetic field dependencies of magnetization, we propose the following micromagnetic model to simulate the experimental data for the Py/Gd bilayer. At low temperatures, the exchange stiffness of a 2-nm thick Gd sublayer and a 2-nm thick sublayer of Py adjacent to the interface [Fig. 1(c)] is relatively low ($A_{Gd\_Int} = A_{Py\_Int} = A_{Int} = 1.5 \times 10^{-7}$ erg/cm). Due to the low stiffness, the magnetizations in these interfacial Gd and Py sublayers begin to twist along the field at $H_{CR}$. A micromagnetic simulation [Fig. 1(d)] shows this twist causes a nonlinear rise of the magnetization when increasing the magnetic field above $H_{CR}$, and the full saturation is not achieved even in high magnetic fields, which is in agreement with the experimental data. The proposed micromagnetic model is also capable of explaining the linear rise of magnetization for the magnetic field below $H_{CR}$ [inset in Fig. 1(b)]. Indeed, if the magnetic order in the top part of the Gd layer is extremely reduced, *e.g.*, its exchange stiffness, $A_{Gd\_Top}$, is of the order of $1.5\times10^{-9}$ erg/cm, then the magnetization in the top Gd sublayer aligns with the magnetic field; whereas the magnetization is directed opposite to the field in the bottom interfacial Gd sublayer. The simulation shows that this transition of spins from parallel to antiparallel-to-magnetic field states is realized through another twist of magnetization within the Gd layer. This twist yields a modest rise of magnetization in low magnetic field [Fig. 1(d)]. The micromagnetic parameters used for the simulation are: $M_{Py} = 810$ emu/cm$^3$, $A_{Py} = 10\times10^{-7}$ erg/cm; $M_{Gd}$ is 1800 emu/cm$^3$ and 1000 emu/cm$^3$ for the bottom (interfacial) and top Gd sublayers, respectively, $A_{Int} = 1.5\times10^{-7}$ erg/cm; $A_{Gd\_Top} = 1.5\times10^{-9}$ erg/cm; exchange stiffness through the Py-Gd interface ($A_{Py-Gd}$) is $1.5\times10^{-7}$ erg/cm.

It was shown that the interfacial exchange coupling between Py and Gd can be controlled by inserting an ultrathin buffer between Gd and TM which blocks intermixing [35]. To determine how the stiffness of the interfacial exchange spring and, hence, $H_{CR}$ is affected by a buffer layer, we inserted a 0.5-nm thick Au buffer between Py and Gd. The first effect of the buffer is that the total magnetic moment at temperatures above 170 K becomes lower than that for the bilayer without Au [Fig. 1(a)]. This means that, due to the reduction of the intermixing, more Gd atoms are ferromagnetic at higher temperatures, therefore, the drop of the total magnetic moment due to antiferromagnetic alignment of Gd and Py is more substantial. Second, $H_{CR}$ for the Py(50 nm)/Au(0.5 nm)/Gd(4 nm) trilayer is 2.4 kOe, which is almost twice as large as $H_{CR}$ for the Py/Gd bilayer (1.3 kOe). From these observations, we draw three conclusions. First, the increase of $H_{CR}$ for the bilayer with the Au layer proves that the interface by itself is not a "weak link" of the coupling, otherwise the presence of the Au layer would lead to a reduction of the coupling and $H_{CR}$. Second, the coupling is defined by the exchange stiffness $A_{Int}$ of the Py and Gd interfacial sublayers, which is strongly affected by the intermixing. Third, the intermixing between Py and Gd and, hence, the effective coupling can be controlled by placing an ultrathin conducting buffer layer between the metals. Our micromagnetic model is capable to fit magnetization curves for both samples, with and without the Au buffer. The only parameter that must be adjusted is $A_{Int}$. $A_{Int} = 2.5\times10^{-7}$ erg/cm provides a good fit of an experimental magnetization curve for the Py(50 nm)/Au(0.5 nm)/Gd(4 nm) film [Fig. 1(d)].

### III. MULTILAYER FILMS

In the micromagnetic model for the bilayers, to account for the change of the exchange stiffness of the Py and Gd near the interface due to proximity effect, the 2-nm-thick sublayers were introduced on each site of the Py/Gd interface. Interdiffusion of Gd and Py was neglected in the model. To determine how the intermixing influences the magnetic properties of an artificial Py/Gd ferrimagnet and to improve the micromagnetic model proposed for explaining the magnetization reversal process in the Py/Gd bilayers, we studied magnetic and magneto-transport properties of the [Py(1 nm)/Gd(1 nm)]$_{25}$ and [Py(2 nm)/Gd(2 nm)]$_{25}$ multilayers over a wide range of temperatures and magnetic fields. The temperature dependencies of the magnetization for these samples measured in small magnetic field (100 Oe) are shown in Fig. 2(a). The [Py(1 nm)/Gd(1 nm)]$_{25}$ film becomes ferromagnetic only at temperatures below 275 K [Fig. 2(a) open dots]. This temperature is lower than the Curie temperatures for bulk Py (850 K) and Gd (292 K). The absence of a clear compensation temperature and the reduced ordering temperature for the thinner Py and Gd layers suggest that the intermixing is a substantial issue for this multilayer. Basically, the intermixing is so significant that, in the first approximation, it can be assumed that the entire [Py(1 nm)/Gd(1 nm)]$_{25}$ film is composed of a PyGd alloy. This observation is confirmed by an X-ray reflectivity measurement (Fig. 3). It is seen that comparatively high superlattice fringes corresponding to the 3.95-nm periodic structure appear only for the film with 2-nm thick layers.

The [Py(2 nm)/Gd(2 nm)]$_{25}$ film demonstrates a more complex, ferrimagnetic-like, temperature dependence of magnetization [Fig. 2(a) solid dots]. Its magnetization is low (70 emu/cm$^3$) at 300 K; most likely, this magnetization is due to thin core parts of the Py layers which are not affected by the intermixing of Py and Gd [inset Fig. 2(a)]. At high temperatures, the magnetization in these core sublayers of Py is aligned along the magnetic field. While the film is cooled down the magnetization of the mixed interfacial regions of Py and Gd (denoted "Mix" further in the text) starts to grow. Importantly, the direction of the magnetization in the Mix sublayers is defined by the magnetic moments of the Gd atoms, which tend to align antiferromagnetically with magnetization in the Py-core sublayers. This leads to a decrease of the total magnetization. At the compensation temperature of 176 K, the magnetic moment of the Py-core sublayers is equal to the magnetic moment of the mixed sublayers, resulting in almost zero (8 emu/cm$^3$) magnetization. Below the compensation temperature, the magnetic moment of the Mix sublayers becomes higher than the moment of the Py-core sublayers, hence, the magnetization is Gd-aligned at these temperatures. Importantly, in contrast to the [Py(1 nm)/Gd(1 nm)]$_{25}$ multilayer, the magnetization begins to rise strongly at temperatures below 75 K. We consider that, similarly to the Py-core sublayers, the core parts of the Gd layers are not affected by the intermixing. The rise of magnetization in these Gd-core sublayers below 75 K causes the upturn of the total magnetization. Taking into account the values of magnetization at 300 K and 10 K as well as assuming that the temperature dependence of the magnetization for the Mix sublayers coincides with that for the [Py(1 nm)/Gd(1 nm)]$_{25}$ film, it was estimated that the effective thicknesses of the Py-core and Gd-core sublayers are about 0.5 nm, wherein the rest of the multilayer is filled with the PyGd alloy [inset in Fig. 2(a)].

Our magnetometry measurements show that for both [Py(2 nm)/Gd(2 nm)]$_{25}$ and [Py(1 nm)/Gd(1 nm)]$_{25}$ films the coercive field does not exceed 10 Oe in the 10–200 K temperature range. This means that the anisotropy of these films is comparable with that for Py, which makes them a good choice for application as a soft artificial ferrimagnet. Noteworthy, the study also showed that the anisotropy becomes significantly higher if the thickness of the Gd layers increases.

The magnetization curves of the [Py(2 nm)/Gd(2 nm)]$_{25}$ multilayer measured at 10 K (black dots) and at the compensation temperature (green dots) are shown in Fig. 2(b). Importantly, even at the compensation temperature, the magnetization of the film is not zero. Furthermore, the magnetization curve does not pass through the origin in zero magnetic field. At the compensation temperature, the magnetization rises linearly in magnetic fields up to 70 kOe, while at 10 K, a complete saturation of magnetization is achieved already in a very small magnetic field (below 50 Oe). We also observe that, at 10 K, the magnetization experiences an abrupt rise at a magnetic field of 16 kOe [Fig. 2(b)], similarly to the one demonstrated by the PyGd bilayers [Fig. 1(b)]. Knowing the estimated thicknesses of the Py-core and Gd-core sublayers [inset in Fig. 2(a)], we modeled the magnetization curves of the [Py(2 nm)/Gd(2 nm)]$_{25}$ multilayer micromagnetically. Fig. 2(c) illustrates simulated the magnetization curves obtained for 10 K (green line) and 176 K (black line). Micromagnetic parameters used for the simulation are $A_{MIX} = A_{Py} = 1.5 \times 10^{-7}$ erg/cm, $M_{Py} = 810$ emu/cm$^3$. For $T = 10$ K, $A_{Gd} = 1.5 \times 10^{-7}$ erg/cm, $M_{MIX} = 861$ emu/cm$^3$, $M_{Gd} = 1600$ emu/cm$^3$. Based on the fact that the Gd-core sublayers gain significant magnetic moment only below 75 K, we assume that the exchange stiffness and magnetization of the Gd-core sublayers is highly suppressed at 176 K, *i.e.*, $A_{Gd} = 0.1 \times 10^{-7}$ erg/cm, $M_{Gd} = M_{MIX} = 116$ emu/cm$^3$. Simulated 10-K magnetization curve shows that, in magnetic field below 15.5 kOe, the magnetizations of the Gd-core and Mix sublayers are pointed along the magnetic field, while the magnetizations in the Py-core sublayers are opposite to the field. When the magnetic field exceeds $H_{CR} = 15.5$ kOe, the magnetization in the very first Py layer, which is affected the least by intermixing, rotates aligning along the magnetic field, resulting in an exchange-spring twist of the magnetization near the bottom surface of the film similarly to the one observed in the PyGd bilayers. The rest of the film preserves antiparallel magnetization alignment along the magnetic field. In order to minimize the total energy at 176 K, when the magnetic moments of the Gd-core/Mix sublayers compensate those of the Py-core sublayers, the corresponding magnetizations tend to align perpendicular to the magnetic field, at the same time, being almost antiparallel to each other (non-collinear configuration). Again, as it was mentioned earlier, the magnetic structure of the very top Py and the very bottom Gd layers is different. Namely, the thicknesses of these Py-core and Gd-core sublayers are 1.25 nm instead 0.5 nm. Basically, it breaks the symmetry of the magnetic structure, and consequently, the magnetizations near the surfaces are at smaller angles with respect to the magnetic field than the magnetization inside the multilayer. To some extent, this configuration is similar to the twisted state predicted by Camley [14, 36-38].

Although the simulated and experimental magnetization curves at 176 K demonstrate qualitatively identical behavior, they do not allow us to determine conclusively the magnetic configuration of the [Py(2 nm)/Gd(2 nm)]$_{25}$ film in the vicinity of the compensation temperature. Electronic transport measurements, on the other hand, are more sensitive to the distribution of the magnetization inside the films and its response to the applied magnetic field. Fig. 4 demonstrates the results of transport measurements for both [Py(1 nm)/Gd(1 nm)]$_{25}$ and [Py(2 nm)/Gd(2 nm)]$_{25}$ films conducted for different magnetic fields and temperatures. First, Fig. 4(a) shows the temperature dependencies of the resistance for the [Py(1 nm)/Gd(1 nm)]$_{25}$ film measured in the 1-kOe (black line) and 100-kOe (blue line) magnetic fields applied longitudinally. These measurements show that the magnetoresistance must change sign at around 40 K. Indeed, Fig. 4(d) illustrates that both longitudinal and transverse magnetoresistances are negative at 50 K while they are positive at 10 K. Second, the same temperature [Fig. 4(b)] and magnetic-field [Fig. 4(c)] dependencies for the [Py(2 nm)/Gd(2 nm)]$_{25}$ film indicate that the magnetoresistance changes sign twice, at around 20 K and, surprisingly, in the vicinity of the compensation temperature. Again, the magnetoresistance is positive at 200 K. In the vicinity of the compensation temperature, the magnetization rotates with respect to the magnetic film and the current, which leads to a change of an anisotropic magnetoresistance. Within the 150–180 K temperature range, a transverse resistance starts to increase while the longitudinal resistance starts to decrease upon increasing the amplitude of magnetic field. Similarly to the [Py(1 nm)/Gd(1 nm)]$_{25}$ film, the magnetoresistance of the [Py(2 nm)/Gd(2 nm)]$_{25}$ film is negative at 50 K and becomes positive again at 10 K.

The anisotropic magnetoresistance is responsible for an interesting step-like change of the resistance observed at 10 K in a 16-kOe magnetic field [Fig. 4(c)]. For the longitudinal resistance, it is an increase, and for the transverse resistance, it is a decrease. These steps are additional evidence that the magnetization of the first Py layer rotates and aligns along the magnetic field. This rotation provides an abrupt rise of the magnetization at 10 K for 16 kOe [Fig. 2(b and c)] as discussed previously. Similar changes of resistance related with nucleation of an in-plane domain-wall have been reported previously for Fe/Gd [29, 30] and Co/Gd [31] systems.

To define the orientation of magnetizations in the [Py(2 nm)/Gd(2 nm)]$_{25}$ multilayer and to disentangle the contribution of the anisotropic magnetoresistance from that of the ordinary and giant magnetoresistances, we measured the angular dependencies of resistance in different magnetic fields. The idea is that, due to anisotropic magnetoresistance, the resistance of the stripe reaches a minimum when the magnetizations in the layers are perpendicular to the direction of the current flow. As an example, the angular dependencies of resistance measured in 10 kOe at 174 and 176 K have minima at 90° and 87°, respectively (Fig. 5). Consequently, the magnetizations in the multilayer are parallel to the 10-kOe magnetic field at 174 and 176 K (collinear configuration). Upon increasing the magnitude of the magnetic field, the minimums begin to shift, indicating that the magnetizations in the Gd-core/Mix and Py-core sublayers rotate with respect to the applied magnetic field. In a 30-kOe magnetic field, the minimum is at 52° at 174 K, and 15° at 176 K. Hence, at 174 K and 176 K, the angles between the magnetization and the 30-kOe magnetic field are at 38° and 75°, respectively. At 70 kOe, both curves in Fig. 5 have minima at around 0°, indicating that the magnetizations are almost perpendicular to the magnetic field. It is worth noting that the shapes of some angular dependencies are not completely sinusoidal, and the amplitude of the angular dependent parts of the curves are different at different temperatures. For example, at 174 K, this amplitude is 23 mΩ in 10 kOe, 7 mΩ in 30 kOe, 19 mΩ in 70 kOe. We assume the reason for this behavior can be a formation of a domain structure. The magnetizations in different domains can be mirrored with respect to the magnetic field. The averaging of the resistance produced by different domains may cause the reduction of the anisotropic magnetoresitance.

To map directions of magnetizations at different temperatures we measured the angular dependence of resistance in different magnetic fields, varying the temperature. Then, based on the positions of the resistance minima, an angle, $α$, between the magnetic field and the axis along which the magnetizations in the Gd-core/Mix and Py-core sublayers are aligned is plotted [Fig. 6(a)]. The same $α(H, T)$ dependence was modeled micromagnetically [Fig. 6(b)]. The most striking and unexpected result provided by the experimental dependence [Fig. 6(a)] is, in a 10-kOe magnetic field, the collinear magnetization configuration is stable even at the compensation temperature, and non-collinear magnetization configurations appear only in higher magnetic fields. The occurrence of a transition from the collinear to non-collinear magnetization configurations even at the compensation temperature cannot be explained by our micromagnetic model presently. It is not clear why the transition does not lead to a horizontal plateau in the magnetization curve at the compensation temperature [Fig. 2(c) open dots]. Secondly, the experimental $α(H, T)$ distribution [Fig. 6(a)] is very narrow. That is, in contrast to the simulations [Fig. 6(b)], the experiment does not show a noncollinear configuration for 70 kOe at temperatures 15 K above or below the compensation temperature.

The unusual phenomena observed in the vicinity of the compensation temperature for the [Py(2 nm)/Gd(2 nm)]$_{25}$ multilayer, namely, the change of magnetoresistance sign, the existence of transition from collinear to noncollinear configurations, and the unexpectedly narrow $α(H, T)$ distribution, can be explained by an inhomogeneous sample structure. We model the magnetic and atomic structures of the film as combination of the Mix, Py-core, Gd-core sublayers, and the thicknesses of these sublayers are the same throughout the film. However, it is possible that accumulating roughness may very well change the amount of the intermixing and may result in the top layers of the [Py(2 nm)/Gd(2 nm)] multilayer more resembling the PyGd alloy. In this case, it is conceivable that the magnetization in this top part is Gd-aligned even above 176 K, while the magnetization in the bottom part of the film is still Py-aligned. Then, above 176 K, one can expect an in-plane domain-wall somewhere inside the film. An increase of the external magnetic field makes this domain-wall narrower, which leads to an increase of the scattering, and consequently, the positive magnetoresistance. Second, the compensation observed at 176 K is not due to the equality of magnetic moments of the Mix/Gd-core and Py-core sublayers. The compensation occurs because the Gd-aligned magnetic moment of strongly intermixed top part of the film becomes equal to the Py-aligned magnetic moment of its bottom part. This yields the $\alpha(H, T)$ distribution to become narrow as it is observed in the experiment. Additionally, such different thickness-dependent intermixing may result in a thickness-dependent variation of the compensation temperature, which in turn would explain the remaining non-zero magnetization even at the experimentally observed compensation temperature.

We believe that the sign-change in magnetoresistance at low temperatures for both [Py(2 nm)/Gd(2 nm)]$_{25}$ and [Py(1 nm)/Gd(1 nm)]$_{25}$ films is caused by an identical mechanism. We do not expect that this sign-change is due to an in-plane domain-wall. First, based on the magnetic-field dependence of the resistance for the [Py(2 nm)/Gd(2 nm)]$_{25}$ film at 10 K [Fig. 4(c)], the magnetoresistance is positive even in magnetic fields lower than the nucleation field of the domain-wall related with rotation of magnetization in the first Py layer. Besides, we do not expect a nucleation of any in-plane domain-wall in the [Py(1 nm)/Gd(1 nm)]$_{25}$ film. A similar change of magnetoresistance sign was observed in a GdNi alloy [39]; where it was speculated that the effect can be related to magnetic polarons induced by Gd.

## IV. EFFECTIVE EXCHANGE STIFFNESS

Analyzing the temperature dependence of magnetization we concluded that the Gd-core sublayers become ferromagnetic only below 75 K [Fig. 2(a)]. We expected that only a short-range exchange exists in the Gd-core sublayers above this temperature. This assumption is then implicitly used for simulating magnetization curve at the compensation temperature when we used $A_{Gd} = 0.1 \times 10^{-7}$ erg/cm. It is peculiar that the magnetic order inside the Gd layers can change so significantly on the scale of 2 nm. To investigate this effect more systematically we fabricated Py(50 nm)/[Py($t$)/Gd($t$)]$_{25}$, $t$ = 1 or 2 nm, films and studied how the effective exchange stiffness of the [Py($t$)/Gd($t$)]$_{25}$ stacks changes upon increasing temperature. The total magnetic moment of the [Py($t$)/Gd($t$)]$_{25}$ stacks is Gd-aligned below 275 K for $t$ = 1 nm and below 176 K for $t$ = 2 nm. Due to the antiferromagnetic coupling, this moment tends to be antiparallel with the magnetic moment of the 50-nm thick Py layer adjacent to the stack. Since the effective exchange stiffness of the stack is expected to be much smaller than that for Py, applying magnetic field leads to an exchange-spring-like rotation of magnetization in the entire stack, and its total magnetic moment aligns along the magnetic field. Fig. 7(a) and (b) illustrate the magnetization distribution over the thickness of the Py(50 nm)/[Py(2 nm)/Gd(2 nm)]$_{25}$ film at 10 K in high and low magnetic fields, respectively. As a model, the entire [Py($t$)/Gd($t$)]$_{25}$ stack can be considered as a homogeneous layer with some effective exchange stiffness, $A_{GdPy}$. This effective layer is antiferromagnetically coupled to the 50-nm thick Py layer. Applying the model for fitting the experimental magnetization curves of the Py(50 nm)/[Py($t$)/Gd($t$)]$_{25}$ film enables estimating $A_{GdPy}$ at different temperatures. The analysis shows that, at 10 K, the Py/Gd stacks of both films can be characterized by the same effective exchange stiffness, $A_{GdPy} = 1.5 \times 10^{-7}$ erg/cm. At 100 K, the effective exchange stiffness of the [Py(1 nm)/Gd(1 nm)]$_{25}$ stack becomes equal to $4 \times 10^{-8}$ erg/cm, while that for the [Py(2 nm)/Gd(2 nm)]$_{25}$ stack is 5 times smaller ($0.8 \times 10^{-9}$ erg/cm). This observation proves that the inner parts of the Gd layers have a lower exchange stiffness and the Py layers are responsible for the enhancement of the magnetic order in the Gd layers.

## V. CONCLUSION

Analysis of the magnetization curves for the Py/Gd bilayers shows that, due to proximity with Py, magnetic order of about 1 nm of the Gd layer adjacent to the Py layer is strongly enhanced. Micromagnetic simulations demonstrate that the magnetization reversal observed in the Py(50 nm)/Gd(4 nm) bilayers in 1.3 kOe is due to an exchange-spring-like twisting of magnetization in the Gd layer. The exchange stiffness of the Gd layer and, hence, parameters of the twist can be controlled by inserting an ultrathin layer of Au between Py and Gd. Using a combination of magnetometry and magnetotransport, we determined the magnetic structures of the Py/Gd multilayers. Based on the reduction of the Curie temperature, it was concluded that the [Py(1 nm)/Gd(1 nm)]$_{25}$ film is mostly composed of a PyGd alloy. Most likely, the [Py(2 nm)/Gd(2 nm)]$_{25}$ film is inhomogeneous over its thickness. Towards the bottom of the film, about 0.5-nm thick inner sublayers of the Py and Gd layers are not affected by the intermixing, while towards the top of the film, the Py and Gd layers are almost completely intermixed. The compensation observed at 176 K is due to an equality of the magnetic moments in the majority of the film. The absence of the local compensation leads to a peculiar behavior of magnetization at the compensation temperature. Namely, using the angular dependencies of resistance, we showed that the noncollinear magnetization configuration is stable only in an unexpectedly narrow range of temperatures, and there is a transition from the collinear to noncollinear configurations in the 10-kOe magnetic field even at the compensation temperature. We believe that the inhomogeneity of the [Py(2 nm)/Gd(2 nm)]$_{25}$ film's magnetic structure is responsible for an in-plane domain-wall in the film at temperatures above 176 K, which leads to the change of magnetoresistance sign at the compensation temperature. Analysis of the magnetization curves for the Py(50 nm)/[Py($t$)/Gd($t$)]$_{25,}$ where $t$ is 1 or 2 nm, multilayers demonstrates that above 75 K the exchange stiffness inside the Gd layers is reduced and magnetic order changes drastically on the scale of 2 nm.

## ACKNOWLEDGEMENTS

Work was supported by the Department of Energy Office of Science, Basic Energy Sciences, Material Sciences and Engineering Division. Pavel N. Lapa also received partial support from Texas A&M University.

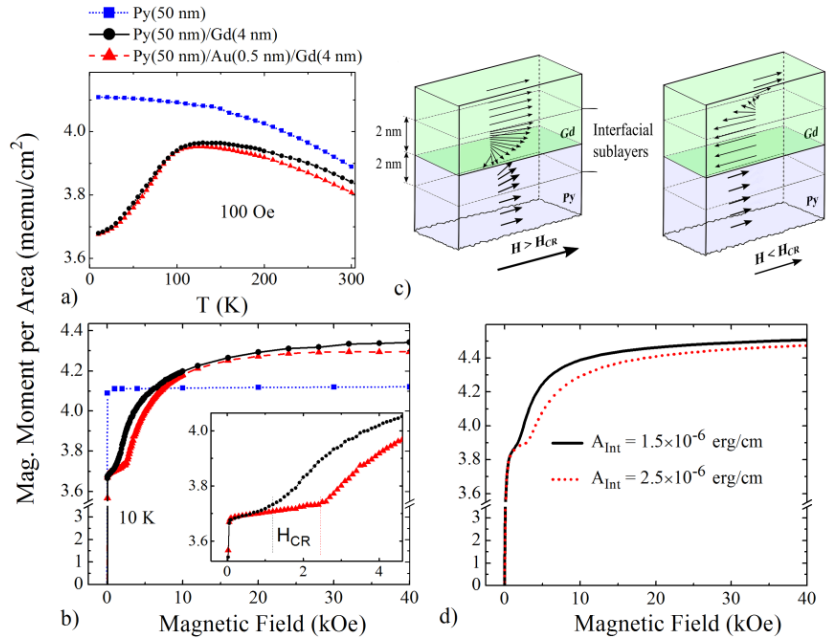

FIG. 1. a) Temperature and b) magnetic field dependencies of the magnetic moment per area for the Py(50 nm), Py(50 nm)/Gd(4 nm), and Py(50 nm)/Au(0.5 nm)/Gd(4 nm) films; c) schematics of exchange-springs in the Py/Gd film above and below $H_{CR}$; d) micromagnetic simulations of the magnetization curves for the Py/Gd film. Exchange stiffness ($A_{Int}$) of the exchange-spring region was varied to simulate reversals for the Py/Gd bilayers with ($A_{Int} = 1.5 \times 10^{-6}$ erg/cm) and without the Au buffer ($A_{Int} = 2.5 \times 10^{-6}$ erg/cm).

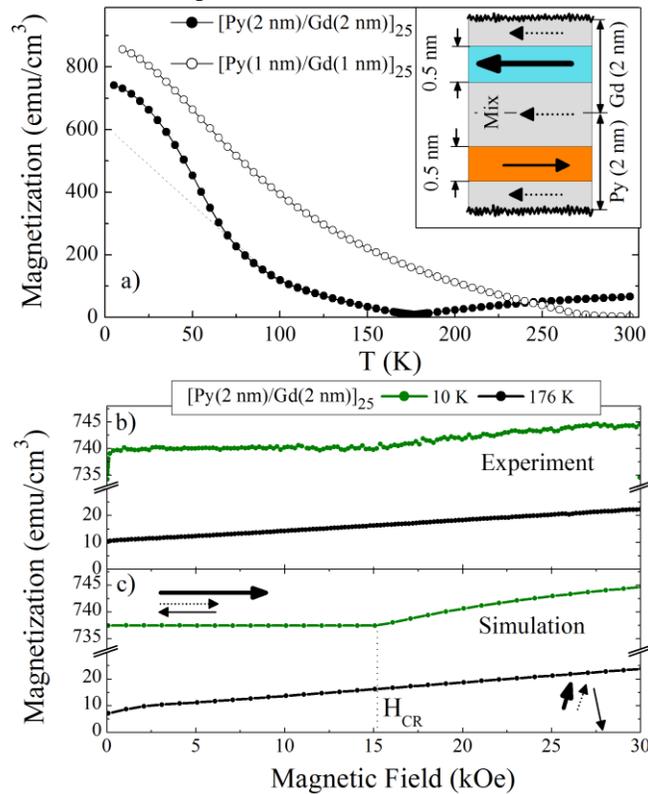

FIG. 2. a) Temperature dependencies of the magnetization for the [Py(1 nm)/Gd(1 nm)]$_{25}$ film (open triangles) and [Py(2 nm)/Gd(2 nm)]$_{25}$ (solid circles) films; a dotted line is drawn to depict the contribution of the Gd-core sublayers to the total magnetization; schematics in the right upper corner illustrates material distribution and positions of the Py-core (orange) and Gd-core (blue) magnetic sublayers in the [Py(2 nm)/Gd(2 nm)]$_{25}$ film; b) Experimental and c) simulated magnetization curves for the [Py(2 nm)/Gd(2 nm)]$_{25}$ film at 10 K (green) and at the compensation temperature (black). Arrows illustrate the direction of magnetizations in the Py-core (thin solid arrow), Gd (thick solid arrow), and Mix (dashed arrow) layers. The insets at the upper-left and bottom-right corners of (c) illustrate magnetization configurations at 10 K and 176 K, respectively.

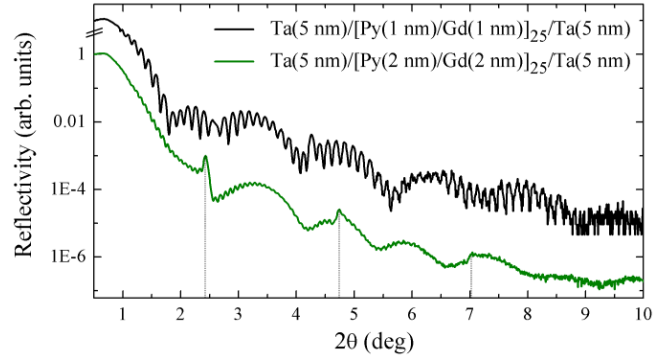

FIG. 3. X-ray reflectivity curves for the [Py(1 nm)/Gd(1 nm)]$_{25}$ (black line) and [Py(2 nm)/Gd(2 nm)]$_{25}$ (green line) films. Short-dashed lines illustrate the positions of the superlattice fringes for the latter film.

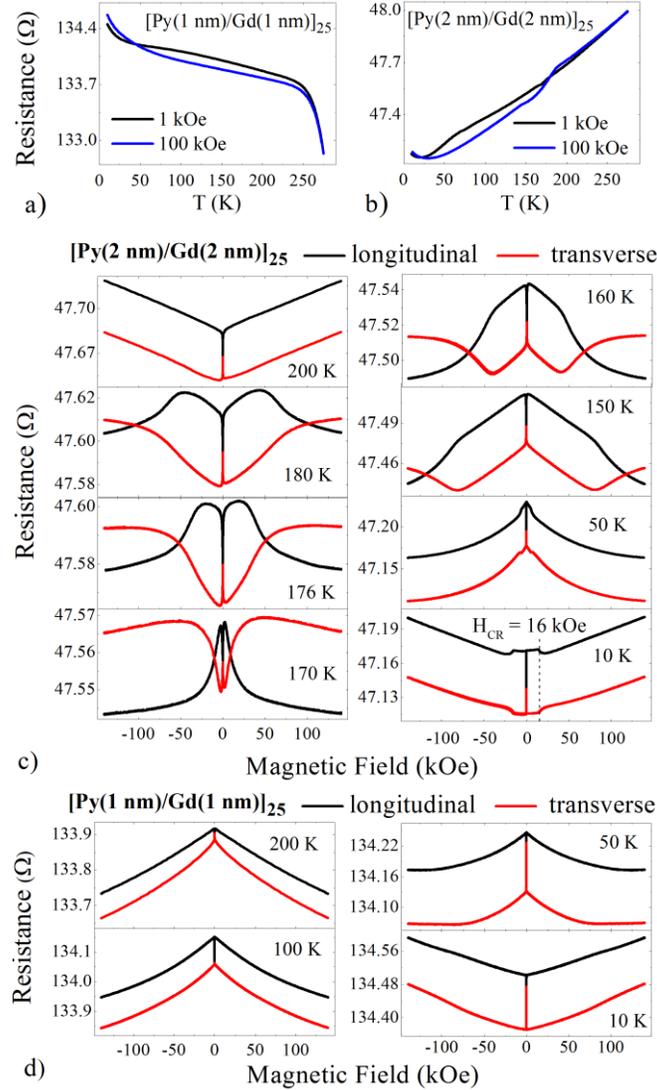

FIG. 4. Temperature dependencies of the resistance measured in 1 kOe (black line) and 100 kOe (blue line) for the (a) [Py(1 nm)/Gd(1 nm)]$_{25}$ and (b) [Py(2 nm)/Gd(2 nm)]$_{25}$ films. Resistance of the (c) [Py(2 nm)/Gd(2 nm)]$_{25}$ and (d) [Py(1 nm)/Gd(1 nm)]$_{25}$ films as a function of longitudinal (black line) and transverse (red line) magnetic field.

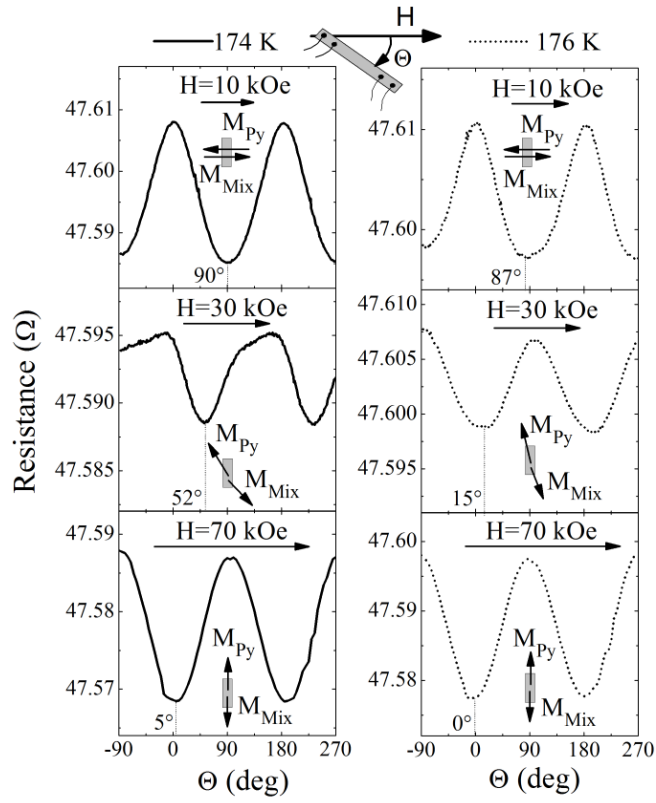

FIG. 5. Angular dependencies of the resistance for the [Py(2 nm)/Gd(2 nm)]$_{25}$ multilayer measured at 174 K and 176 K in 10-, 30-, 70-kOe magnetic fields. The top central insert demonstrates the direction of the stripe rotation. A zero-angle corresponds to a position where the long edge of the stripe is along the magnetic field. The small insets near each curve depict the position of the stripe at 90° and orientation of magnetizations in the core-Py (M$_{Py}$) and Mix (M$_{Mix}$) sublayers.

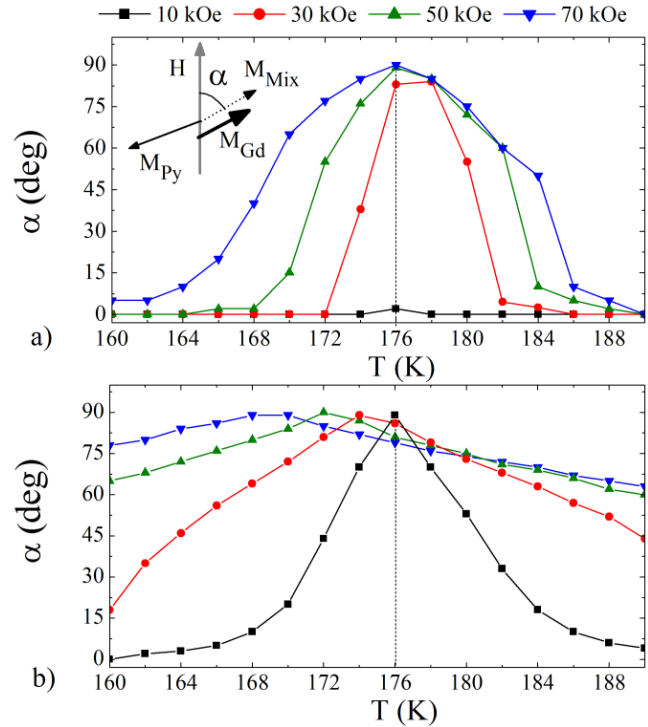

FIG. 6. a) Experimental and b) simulated temperature dependencies of the angle between magnetic field and a line along which the magnetizations in the [Py(2 nm)/Gd(2 nm)]$_{25}$ film are predominantly aligned.

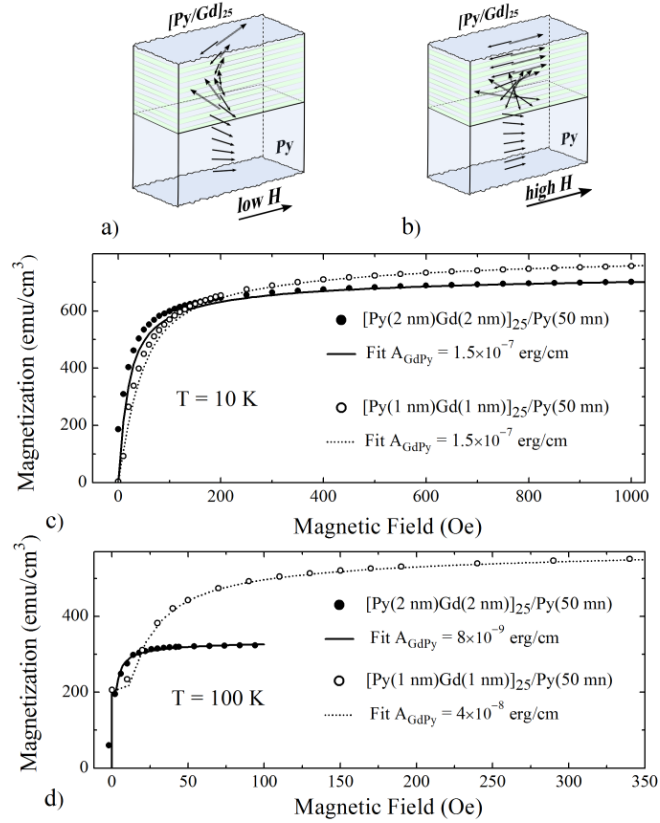

FIG. 7. The in-plane domain-walls in Py(50 nm)/[Py(2 nm)/Gd(2 nm)]$_{25}$ in high (a) and low (b) magnetic fields. Magnetization curves of Py(50 nm)/[Py(2 nm)/Gd(2 nm)]$_{25}$ (solid dot – experimental data, solid lines – fits) and Py(50 nm)/[Py(1 nm)/Gd(1 nm)]$_{25}$ (open dots – experimental data, short-dashed line – fits) at 10 K (c), and 100 K (d).

**REFERENCES**

1. R. Wu and A. J. Freeman, Journal of Magnetism and Magnetic Materials **99**, 81-84 (1991).
2. A. Heys, P. E. Donovan, A. K. Petford-Long and R. Cywinski, Journal of Magnetism and Magnetic Materials **131**, 265-272 (1994).
3. A. Heys and P. E. Donovan, Journal of Magnetism and Magnetic Materials **126**, 326-328 (1993).
4. A. Barth, F. Treubel, M. Marszałek, W. Evenson, O. Hellwig, C. Borschel, M. Albrecht and G. Schatz, Journal of Physics: Condensed Matter **20**, 395232 (2008).
5. M. Taborelli, R. Allenspach, G. Boffa and M. Landolt, Physical Review Letters **56**, 2869-2872 (1986).
6. D. Haskel, G. Srajer, J. C. Lang, J. Pollmann, C. S. Nelson, J. S. Jiang and S. D. Bader, Physical Review Letters **87**, 207201 (2001).
7. J. W. A. Robinson, F. Chiodi, M. Egilmez, G. B. Halász and M. G. Blamire, Scientific Reports **2**, 699 (2012).
8. T. Katayama, M. Hirano, Y. Koizumi, K. Kawanishi and T. Tsushima, IEEE Transactions on Magnetics **13**, 1603-1605 (1977).
9. M. Nawate, K. Doi and S. Honda, Journal of Magnetism and Magnetic Materials **126**, 279-281 (1993).
10. T. Eimüller, R. Kalchgruber, P. Fischer, G. Schütz, P. Guttmann, G. Schmahl, M. Köhler, K. Prügl, M. Scholz, F. Bammes and G. Bayreuther, Journal of Applied Physics **87**, 6478-6480 (2000).
11. S. Tomonori, O. Kouji, O. Kenshou, O. Yutaka, M. Shunsuke and S. Yoshifumi, Japanese Journal of Applied Physics **13**, 201 (1974).
12. J. C. T. Lee, J. J. Chess, S. A. Montoya, X. Shi, N. Tamura, S. K. Mishra, P. Fischer, B. J. McMorran, S. K. Sinha, E. E. Fullerton, S. D. Kevan and S. Roy, Applied Physics Letters **109**, 022402 (2016).
13. S. A. Montoya, S. Couture, J. J. Chess, J. C. T. Lee, N. Kent, D. Henze, S. K. Sinha, M.-Y. Im, S. D. Kevan, P. Fischer, B. J. McMorran, V. Lomakin, S. Roy and E. E. Fullerton, arXiv:1608.01368.
14. R. E. Camley and D. R. Tilley, Physical Review B **37**, 3413-3421 (1988).
15. C. D. Stanciu, F. Hansteen, A. V. Kimel, A. Kirilyuk, A. Tsukamoto, A. Itoh and T. Rasing, Physical Review Letters **99**, 047601 (2007).

16. S. Mangin, M. Gottwald, C. H. Lambert, D. Steil, V. Uhlíř, L. Pang, M. Hehn, S. Alebrand, M. Cinchetti, G. Malinowski, Y. Fainman, M. Aeschlimann and E. E. Fullerton, Nat Mater **13**, 286-292 (2014).
17. T. A. Ostler, J. Barker, R. F. L. Evans, R. W. Chantrell, U. Atxitia, O. Chubykalo-Fesenko, S. El Moussaoui, L. Le Guyader, E. Mengotti, L. J. Heyderman, F. Nolting, A. Tsukamoto, A. Itoh, D. Afanasiev, B. A. Ivanov, A. M. Kalashnikova, K. Vahaplar, J. Mentink, A. Kirilyuk, T. Rasing and A. V. Kimel, Nature Communications **3**, 666 (2012).
18. K. Andrei, V. K. Alexey and R. Theo, Reports on Progress in Physics **76**, 026501 (2013).
19. C. Xu, T. A. Ostler and R. W. Chantrell, Physical Review B **93**, 054302 (2016).
20. C.-H. Lambert, S. Mangin, B. S. D. C. S. Varaprasad, Y. K. Takahashi, M. Hehn, M. Cinchetti, G. Malinowski, K. Hono, Y. Fainman, M. Aeschlimann and E. E. Fullerton, Science **345**, 1337-1340 (2014).
21. R. Morales, J. I. Martín and J. M. Alameda, Physical Review B **70**, 174440 (2004).
22. C. Blanco-Roldán, Y. Choi, C. Quirós, S. M. Valvidares, R. Zarate, M. Vélez, J. M. Alameda, D. Haskel and J. I. Martín, Physical Review B **92**, 224433 (2015).
23. S. Honda, M. Nawate and I. Sakamoto, Journal of Applied Physics **79**, 365-372 (1996).
24. R. Ranchal, C. Aroca, M. Maicas and E. López, Journal of Applied Physics **102**, 053904 (2007).
25. R. Ranchal, C. Aroca and E. López, Journal of Applied Physics **100**, 103903 (2006).
26. R. Ranchal, Y. Choi, M. Romera, J. W. Freeland, J. L. Prieto and D. Haskel, Physical Review B **85**, 024403 (2012).
27. D. Larbalestier, A. Gurevich, D. M. Feldmann and A. Polyanskii, Nature **414**, 368-377 (2001).
28. L. T. Baczewski, R. Kalinowski and A. Wawro, Journal of Magnetism and Magnetic Materials **177**, 1305-1307 (1998).
29. M. Vaezzadeh, B. George and G. Marchal, Physical Review B **50**, 6113-6118 (1994).
30. K. Takanashi, Y. Kamiguchi, H. Fujimori and H. Motokawa, J. Phys. Soc. Jpn. **61**, 3721-3731 (1992).
31. J. Colino, J. P. Andrés, J. M. Riveiro, J. L. Martínez, C. Prieto and J. L. Sacedón, Physical Review B **60**, 6678-6684 (1999).
32. K. Takanashi, H. Kurokawa and H. Fujimori, Applied Physics Letters **63**, 1585-1587 (1993).
33. D. LaGraffe, P. A. Dowben and M. Onellion, Journal of Vacuum Science & Technology A **8**, 2738-2742 (1990).
34. T. McGuire and R. Gambino, IEEE Transactions on Magnetics **14**, 838-840 (1978).
35. R. Ranchal, C. Aroca, M. Sánchez, P. Sánchez and E. López, Applied Physics A **82**, 697-701 (2006).
36. R. E. Camley, Physical Review B **35**, 3608-3611 (1987).
37. Y. Jian, Solid State Communications **74**, 1221-1224 (1990).
38. R. E. Camley, Physical Review B **39**, 12316-12319 (1989).
39. R. Mallik, E. V. Sampathkumaran, P. L. Paulose and V. Nagarajan, Physical Review B **55**, R8650-R8653 (1997).